\documentclass{ptephy}

\usepackage{amsmath} 
\usepackage{amsthm} 
\usepackage{cite} 
\usepackage{graphicx} 
\usepackage{algorithmic} 
\usepackage{subfig} 
\usepackage{url} 
\usepackage{latexsym}
\usepackage{natbib}
\usepackage{bm}
\bibliography{sample}

\begin{document}
\title{Cosmological models with the energy density of random fluctuations and the Hubble-constant problem}
\author{\name{Kenji Tomita}{\ast}}
\address{\affil{}{Yukawa Institute for Theoretical Physics, Kyoto University, Kyoto 606-8502,
 Japan}
\rm{\email{ketomita@ybb.ne.jp}}}

\begin{abstract}
 The fluctuation energy is derived from adiabatic random 
 fluctuations due to the second-order perturbation theory, and the
evolutionary relation
 for it is expressed in the form of $\rho_f = \rho_f (\rho)$, where
$\rho$ and $ \rho_f$  are the densities of ordinary dust and the 
fluctuation energy, respectively. The pressureless matter 
as a constituent of the universe at the later stage is assumed to 
consist of ordinary dust and the fluctuation energy. 
Next,  cosmological models including 
the fluctuation energy as a kind of dark matter are derived using the above
relation, and it is found that the
 Hubble parameter and the other model parameters in the derived models 
 can be consistent  with the recent observational values. Moreover, the
 perturbations of $\rho$ and $\rho_f$ are studied. 

\end{abstract}

\maketitle

\section{Introduction}
At the later stage of the universe, the main constituent is 
considered to be a pressureless matter consisting of ordinary dust. 
It is well known that the universe has random 
fluctuations in its density which were caused by  quantum 
fluctuations  at the early stage\citep{wein,ll,dodel,tsuji,bbks}, and their 
amplitude and spectrum have been studied through precise mesurements 
of fluctuations in the cosmic microwave background radiation (CMB) 
 by WMAP\citep{wmap} and Planck\citep{planck1,planck2} collaborations.
  However the mean energy density corresponding to
the fluctuations has not been derived, and so their dynamical influence 
on the universe has also not been clarified yet.

 In a previous paper\citep{tompre},  we tried to derive the energy 
 density of random
 fluctuations using the general-relativistic second-order nonlinear
 perturbation theory\citep{tom,tomold}, in which the random density
  fluctuations are given 
 as the first-order density perturbations with the specified spectrum, and
 the homogeneous energy density $\rho_f$ was derived as the
 (spatially) averaged value of the second-order density perturbations.
 Moreover, the corresponding second-order metric perturbations and
 its spatial average were also derived. 
By adding the contribution of second-order homogeneous 
perturbations to the background model parameters, we renormalized 
the model parameters
from the background ones to modified ones. As a result of this
 procedure, we found the possibility of solving the Hubble-constant problem, 
 in which  the contradiction between the measured Hubble constant and
 the background Hubble constant was
  shown\citep{planck1,planck2,h1,h2,h21,h3,h4,decay}.  In the previous paper, 
  it was found that the renormalized Hubble constant can become 
  nearly equal to the measured Hubble constants.
 
 In this paper, we treat the fluctuation energy as a kind of dark matter 
 and construct cosmological models involving it as part of the
 constituent.  In Sect.2 we express  the fluctuation energy density
 $\rho_f$ as a function of the ordinary dust density $\rho$, using
 the result of calculations in the second-order perturbation theory
 in the basic background models.
 In Sect.3, we derive cosmological flat models including  
pressureless matter whose density is the sum of the densities of
ordinary dust ($\rho$) and  the fluctuation energy ($\rho_f$).
The revised model parameters in these models are compared with 
those in the basic models without the fluctuation energy.
In Sect. 4, we discuss the perturbations in the models with the fluctuation
energy. In Sect.5, we give some concluding remarks. In Appendix A, the
formula of the fluctuation energy is shown.
\bigskip
 
\section{Evolutionary relation for the fluctuation energy}
First, to derive the fluctuation energy,  we assume two basic background
 models (Model 1 and Model 2) with
\begin{equation}
  \label{eq:1}
(\Omega_M^b, \Omega_\Lambda^b) = (0.22, 0.78) \ {\rm and} \ 
H_0^b = 67.3 \ {\rm km \ s^{-1} Mpc^{-1}} \quad {\rm for \ Model \ 1} ,
\end{equation}
and 
\begin{equation}
  \label{eq:1a}
(\Omega_M^b, \Omega_\Lambda^b) = (0.24, 0.76) \ {\rm and} \ 
H_0^b = 67.3 \ {\rm km \ s^{-1} Mpc^{-1}} \quad {\rm for \ Model \ 2} ,
\end{equation}
where 
\begin{equation}
  \label{eq:2}
\Omega_M^b  = \frac{8\pi G \rho^b_0}{3(H^b_0)^2} = \frac{1}{3}
 \frac{\rho^b_0}{(H^b_0)^2}  \quad {\rm{and}} \quad \Omega^b_{\Lambda} =
  \frac{\Lambda c^2}{3(H^b_0)^2} = \frac{1}{3}  \frac{\Lambda}{(H^b_0)^2},
\end{equation}
$\rho^b$ is the density of ordinary dust in the basic background models, 
$H^b_0$ is the Hubble parameter $H^b$ at the present epoch $t_0^b$,
 and \ $8\pi G = c = 1$. In the previous paper\citep{tompre}, only Model 1
  was taken as the background model.
 Here we also consider Model 2 for reference.
The Hubble parameter $H^b$ satisfies
\begin{equation}
  \label{eq:3}
(H^b)^2 = \frac{1}{3} (\rho^b + \Lambda).
\end{equation}

Using the transfer function (BBKS) for cold dark matter adiabatic
fluctuations\citep{bbks}, we derived the second-order density perturbations 
$\delta_2 \rho$, and the spatial average $\langle \delta_2 \rho \rangle$
as a function of the cosmic time $t^b$
in the previous paper. The formula for  $\langle \delta_2 \rho
 \rangle$ is shown in Appendix A.
The latter is represented here as the fluctuation energy $\rho_f \ (\equiv 
\langle \delta_2 \rho \rangle)$.   In this paper, we eliminate $t^b$ from
 $\rho_f$ and $\rho^b$, and  represent $\rho_f$ as the evolutionary 
function ($\rho_f (\rho^b)$) of $\rho^b$. 
Moreover,  the ratio of their values is  expressed  as
\begin{equation}
  \label{eq:4}
\beta (\rho^b) \equiv \rho_f (\rho^b)/\rho^b.
\end{equation}
The value $\beta$ at $\rho^b \rightarrow \infty$ vanishes and the present
values are
\begin{equation}
  \label{eq:5}
\beta (\rho_0^b) = 0.552 \ \ {\rm and} \ \ 0.685
\end{equation}
for Models 1 and 2, respectively.
Their numerical values in Models 1 and 2 are
 shown as  functions of $\rho^b$ in Figs.1 and 2, respectively.
 
The functional relation $\beta (x)$ is expressed approximately using 
two analytic functions in Models 1 and 2 in terms of $u \equiv 
x/[3(H^b_0)^2]$ as follows. 

\noindent Model 1:
\begin{equation}
  \label{eq:6}
\beta (x) = 0.292 (1/u)^{0.400} [1+ 0.065/u -0.0137/u^2]  
\end{equation}
for  \ $(0.22)^{-1} \geq  1/u \geq 0.982 \quad (1 \geq a^b \geq 0.6)$, and
\begin{equation}
  \label{eq:7}
\beta (x) = 0.383 (1/u)^{0.665} [1- 0.261/u + 0.0627/u^2]  
\end{equation}
for  \ $0.982 \geq 1/u \geq 0 \quad (0.6 \geq a^b \geq 0)$,
where $a^b$ is the scale-factor with the present value $a^b_0 = 1$.

\noindent Model 2:
\begin{equation}
  \label{eq:6a}
\beta (x) = 0.362 (1/u)^{0.400} [1+ 0.0884/u -0.0166/u^2]  
\end{equation}
for  \ $(0.24)^{-1} \geq 1/u \geq 0.9 \quad (1 \geq a^b \geq 0.6)$, and
\begin{equation}
  \label{eq:7a}
\beta (x) = 0.459 (1/u)^{0.665} [1+ 0.074/u -0.251/u^2]  
\end{equation}
for  \ $0.900 \geq 1/u \geq 0 \quad (0.6 \geq a^b \geq 0)$.

\bigskip
\bigskip
\begin{figure}
\caption{\label{fig:1}  $\beta$  is expressed as 
a function of $1/u$ in Model 1. The ordinate is $\beta \ (\equiv \rho_f/\rho^b)$  and $u \equiv \rho^b/[3(H^b_0)^2]$.}
\centerline{\includegraphics[width=10cm]{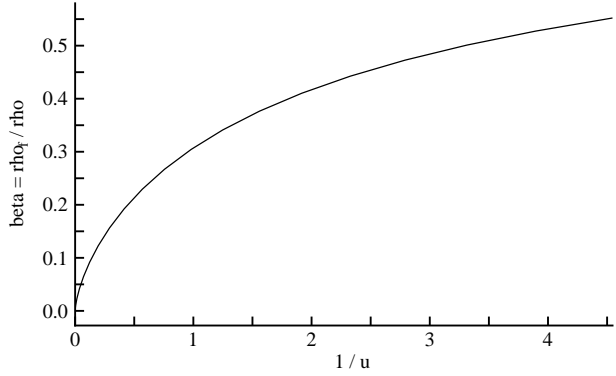}}
\end{figure}
\begin{figure}
\caption{\label{fig:2}  $\beta$  is expressed as 
a function of $1/u$ in Model 2. The ordinate is $\beta \ (\equiv \rho_f/\rho^b)$  and $u \equiv \rho^b/[3(H^b_0)^2]$. }
\centerline{\includegraphics[width=10cm]{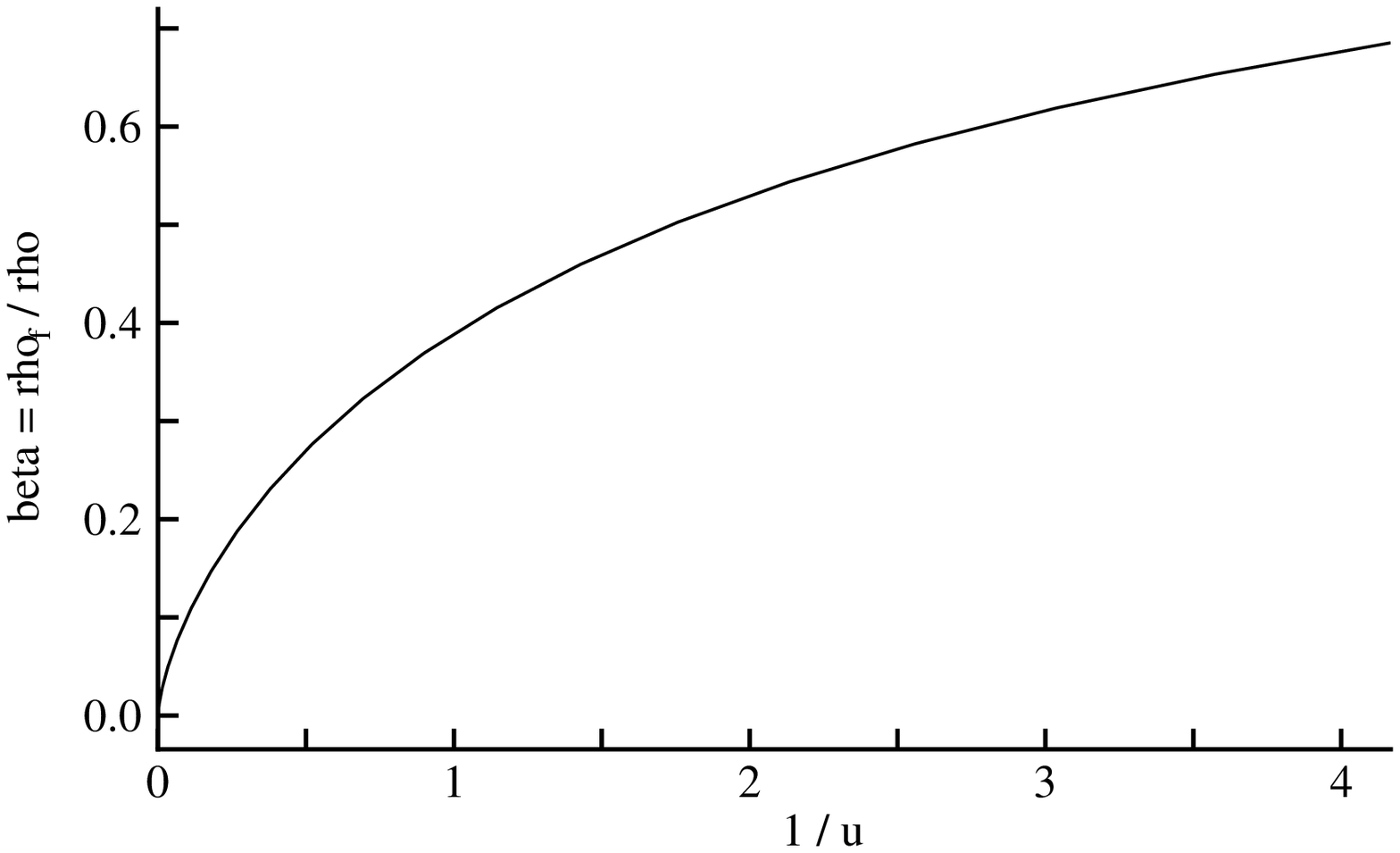}}
\end{figure}
 
\bigskip

\section{Cosmological models with the fluctuation energy and the model
 parameters}
To derive a spatially flat model with the fluctuation energy, 
we consider the line element 
\begin{equation}
  \label{eq:8}
ds^2 = g_{\mu \nu} dx^\mu dy^\nu = a^2 (\eta) [-d\eta^2 + \delta_{ij} 
dx^i dx^j ],
\end{equation}
where the Greek and Roman letters denote $0, 1, 2, 3$ and $1, 2, 3$,
 respectively. The conformal time $\eta (= x^0)$ is related to the cosmic
  time $t$ by $dt = a(\eta) d\eta$.

In this paper, the fluctuation energy is regarded as a kind of dark matter, 
and is assumed to move together with ordinary dust. 
  Then  the velocity vector and energy-momentum tensor of 
pressureless matter are expressed in comoving coordinates as
\begin{equation}
  \label{eq:9}
u^0 = 1/a,  \quad u^i = 0
\end{equation}
and
\begin{equation}
  \label{eq:10}
T^0_0 = -\rho_T, \quad T^0_i = 0, \quad T^i_j = 0
\end{equation}
with $\rho_T \equiv \rho + \rho_f$, where $\rho_T, \rho,$ and  $\rho_f$ 
are the total density of pressureless matter, the ordinary dust density,
and the fluctuation energy density, respectively, and we assume 
\begin{equation}
  \label{eq:10a}
\rho_f = \rho \ \beta (\rho)
\end{equation}
as the approximate equation of state for the fluctuation energy, where
the function $\beta (\rho)$ is specified by Eq.(\ref{eq:4}) with Figs. 1 
and 2, and Eqs.(\ref{eq:6}) $\sim$ (\ref{eq:7a}).

From the Einstein equations, we obtain    
\begin{equation}
  \label{eq:11}
 \rho_T a^2 = 3(a'/a)^2 - \Lambda a^2,
\end{equation}
and the energy-momentum conservation ($T^{\mu\nu}_{;\nu} = 0$) 
gives the relation 
\begin{equation}
  \label{eq:12}
\rho_T a^3 = \rho_T (t_0),
\end{equation}
where $a = 1$ at the present epoch ($t = t_0$) and a prime denotes 
$\partial/\partial \eta$.
In the previous paper\citep{tompre}, the renormalization of the Hubble
 constant
was done using the spatial average of the second-order metric 
perturbation. In this paper, the Hubble parameter is derived only through 
considering the fluctuation energy $\rho_f$ as the part of the total energy. 
Then the Hubble parameter $H (\equiv \dot{a}/a = a'/a^2)$ satisfies
\begin{equation}
  \label{eq:13}
H^2 = \frac{1}{3}(\rho_T + \Lambda) =  \frac{1}{3}(\rho + \rho_f + \Lambda)
\end{equation}
and we have the relations for the model parameters
\begin{equation}
  \label{eq:15}
\Omega_M \equiv \Omega_{Md} + \Omega_{Mf},
\end{equation}
and
\begin{equation}
  \label{eq:16}
\Omega_{Md} \equiv \frac{1}{3} \frac{\rho (t_0)}{(H_0)^2}, \quad
 \Omega_{Mf}  \equiv \frac{1}{3} \frac{\rho _f (t_0)}{(H_0)^2}, 
\end{equation}
where $H_0$ is $H$ at the present epoch ($t_0$).
This model reduces to the basic background models in Sect. 2 
in the limit  $a \rightarrow 0$, because $\rho_f/\rho \rightarrow 0$.

From Eqs.(\ref{eq:11}) and (\ref{eq:15}), the equation for $a$ is 
\begin{equation}
  \label{eq:17}
a' = H_0 [a (\Omega_M + \Omega_\Lambda a^3)]^{1/2},
\end{equation}
and  $a(t)$ is determined by specifying $\Omega_M, 
\Omega_\Lambda$, and $H_0$, and solving this equation.
 
Now let us derive the model parameters  $(\Omega_M, 
\Omega_\Lambda, H_0) $ in the present model 
as the function of $(\Omega_M^b, \Omega_\Lambda^b, H_0^b) $ in
the basic models. Here the Hubble parameters are represented by $H$ 
and $H^b$ at epochs with scale factors
$a$ and $a^b$, respectively,  and  their ratio $\alpha$ is expressed as
\begin{equation}
  \label{eq:18}
\alpha^2 \equiv (H/H^b)^2 = \frac{(1+\beta)\rho +\Lambda}{\rho^b +\Lambda}
\end{equation}
using Eqs. (\ref{eq:3}) and (\ref{eq:13}). This equation is rewritten  as
\begin{equation}
  \label{eq:19}
\rho/\rho^b = [(\alpha^2 -1) \Lambda/\rho^b + \alpha^2]/(1 + \beta). 
\end{equation}
At the present epoch with $a  =  a^b  = 1$, we have
\begin{equation}
  \label{eq:20}
(\alpha_0)^2 = \Omega_\Lambda^b + (1+\beta_0) \Omega_M^b
 (\rho/\rho^b)_0\
\end{equation}
or
\begin{equation}
  \label{eq:21}
(\rho/\rho^b)_0 = X/(1+\beta_0),
\end{equation}
where $(\alpha_0, \beta_0)$ is the present counterpart of $(\alpha, \beta)$
and 
\begin{equation}
  \label{eq:22}
X \equiv [(\alpha_0)^2 -1]\Omega_\Lambda^b/\Omega_M^b + (\alpha_0)^2.
\end{equation}

Here we express $\Omega_{Md}, \Omega_{Mf},$ and $\Omega_\Lambda$
in terms of $\Omega_M^b$ and $\Omega_\Lambda^b$. 
Using Eq.(\ref{eq:20}), we obtain
\begin{equation}
  \label{eq:23}
(\Omega_{Md}, \Omega_{Mf}, \Omega_\Lambda) = \frac{1}{3(H_0)^2}
(\rho_0, \rho_{f0},\Lambda)
= \frac{(\rho_0, \rho_{f0},\Lambda)}{3(\alpha_0)^2 (H_0^b)^2}.  
\end{equation}
Using Eq.(\ref{eq:21}), moreover, we obtain
\begin{equation}
  \label{eq:23a}
(\Omega_{Md}, \Omega_{Mf}, \Omega_\Lambda) = \frac{\Omega_M^b
}{(\alpha_0)^2} (\frac{X}{1+\beta_0}, \frac{\beta_0 X}{1+\beta_0},
 \frac{\Omega_\Lambda^b}{\Omega_M^b}).
\end{equation}
For the density parameter of the pressureless matter $\Omega_M 
\ (\equiv \Omega_{Md} + \Omega_{Mf})$, we have
\begin{equation}
  \label{eq:24}
(\Omega_M, \Omega_\Lambda) = \frac{\Omega_M^b}{(\alpha_0)^2}
 (X, \Omega_\Lambda^b/\Omega_M^b),
\end{equation}
and for ordinary dust, we have
\begin{equation}
  \label{eq:25}
(\Omega_{Md}, \Omega_{Mf} + \Omega_\Lambda) =  \frac{\Omega_M^b
}{(\alpha_0)^2} (\frac{X}{1+\beta_0}, \frac{\beta_0}{1+\beta_0} X
 +\Omega_\Lambda^b/\Omega_M^b).
\end{equation}

Here we consider the correspondence between the ordinary dust 
density in the model with 
$\rho_f \neq 0$ and that in the basic model ($\rho_f = 0$), 
so that we may clarify the additional effect of the fluctuation energy.
First we take the correspondence in which the present densities
of  ordinary dust are equal, i.e.,
\begin{equation}
  \label{eq:26}
(\rho/\rho^b)_0 = 1 .
\end{equation}
  Then we obtain $X = 1 + \beta_0$ 
and
\begin{equation}
  \label{eq:27}
(\alpha_0)^2 = \Omega_M^b (1+ \beta_0) + \Omega_\Lambda^b
\end{equation}
from Eq.(\ref{eq:20}). For inserting $\beta_0 \equiv \beta 
(\rho_0)$ and the model parameters
of the two basic models, therefore, we obtain
\begin{equation}
  \label{eq:28}
   \begin{split}
\alpha_0 &= 1.059, \cr
 (\Omega_M, \Omega_\Lambda) &= (0.305, 0.695) 
\ {\rm and} \ H_0 = 71.3 \ {\rm km \ s^{-1} Mpc^{-1}}
 \end{split}
\end{equation}
for Model 1, and
\begin{equation}
  \label{eq:28a}
   \begin{split}
\alpha_0 &= 1.079, \cr
 (\Omega_M, \Omega_\Lambda) &= (0.347, 0.653) 
\ {\rm and} \ H_0 = 72.6 \ {\rm km \ s^{-1} Mpc^{-1}}
 \end{split}
\end{equation}
for Model 2.

For the present ordinary dust density ratio $(\rho/\rho^b)_0$ which
 is not equal to $1$, we have
$X = (\rho/\rho^b)_0 (1 + \beta_0)$ from Eq.(\ref{eq:21}), and,
using Eq.(\ref{eq:20}) for $\alpha_0$, we obtain the following parameters 
for the model parameters of the two basic models and  several values 
of $(\rho/\rho^b)_0$ :
\begin{equation}
  \label{eq:31}
    \begin{split}
(\rho/\rho^b)_0 &= 1.181, \quad \alpha_0 = 1.088, \cr
(\Omega_M , \Omega_\Lambda) &= (0.341, 0.659), 
\ {\rm and} \ 
H_0 = 73.2 \ {\rm km \ s^{-1} Mpc^{-1}}
  \end{split}
\end{equation}
for Model 1, and
\begin{equation}
  \label{eq:32}
   \begin{split}
(\rho/\rho^b)_0 &= 1.110, \quad \alpha_0 = 1.099, \cr
 (\Omega_M , \Omega_\Lambda) &= (0.371, 0.629), 
\ {\rm and} \ 
H_0 = 74.0 \ {\rm km \ s^{-1} Mpc^{-1}},
 \end{split}
\end{equation}
and
\begin{equation}
  \label{eq:33}
   \begin{split}
(\rho/\rho^b)_0 &= 1.276, \quad \alpha_0 = 1.130, \cr
 (\Omega_M , \Omega_\Lambda) &= (0.404, 0.596), 
\ {\rm and} \ 
H_0 = 76.0 \ {\rm km \ s^{-1} Mpc^{-1}}
 \end{split}
\end{equation}
for Model 2.

Thus, we obtained  model parameters in the models with fluctuation energy 
by specifying the basic model parameters and $(\rho/\rho^b)_0$ for 
their correspondence. 
The above model parameters with the fluctuation energy are comparable 
with the observed ones.\citep{planck1,planck2,h1,h2,h3,h4,decay} 
Those with $(\rho/\rho^b)_0 = 1$ in Model 2 and $(\rho/\rho^b)_0 
= 1.181$ in Model 1 are  near to the observed ones with $H_0 = 73 \sim
 74 \ {\rm km \ s^{-1} Mpc^{-1}}$.

\medskip

Next let us study the behaviors of models in the past in 
comparison with the basic models.  Here $\alpha  \ (\equiv H/H^b)$ 
is obtained from Eq. (\ref{eq:18}) as 
\begin{equation}
  \label{eq:34}
\alpha^2 = \frac{(\rho_T)_0/a^3 +\Lambda}{(\rho^b/\rho_T) (\rho_T)_0/a^3 
+ \Lambda} = \frac{(1+\beta_0)(\rho/\rho^b)_0 \Omega_M^b
 +\Omega^b_\Lambda a^3}
{\frac{1+\beta_0}{1+\beta} \frac{(\rho/\rho^b)_0}
{\rho/\rho^b}\Omega_M^b +\Omega^b_\Lambda a^3},
\end{equation}
where $\beta (\rho)$ is given by Eqs. (\ref{eq:6}) - (\ref{eq:7a}), and 
\begin{equation}
  \label{eq:35}
a^3 = \rho_T(t_0)/\rho_T = (\rho_0/\rho) \frac{1 + \beta_0}{1+ \beta(\rho)}.
\end{equation}

To evaluate $\alpha$ in the past for $(\rho/\rho^b)_0 = 1$, 
we take  the correspondence between  $a$ and $a^b$, in such a way that 
$\rho/\rho^b = 1$ also in the past. Then we have
\begin{equation}
  \label{eq:36}
\alpha^2  = \frac{(1+\beta_0) \Omega_M^b
 +\Omega^b_\Lambda a^3}
{\frac{1+\beta_0}{1+\beta} \Omega_M^b +\Omega^b_\Lambda a^3},
\end{equation}
so that $\beta \rightarrow 0$  and  $\alpha \rightarrow 1$  for
$a \rightarrow 0$.

To evaluate $\alpha$ in the past for $(\rho/\rho^b)_0 
\neq 1$, we take the correspondence in such a way that 
\begin{equation}
  \label{eq:37}
\rho/\rho^b = [(\rho/\rho^b)_0 -1] (\rho^b_0/\rho^b) +1.
\end{equation}
Then we find that $ \rho/\rho^b \rightarrow 1$ 
and $\beta \rightarrow 0$ for $a \rightarrow 0$, and 
 from Eq.(\ref{eq:34}) that $\alpha \rightarrow 1$ for $a \rightarrow 0$.

The $a$ dependences of $1/u, \beta$ and 
$\alpha$ in the case of $(\rho/\rho^b)_0 = 1$ for Model 2 (with the
model parameter (\ref{eq:28a})) are  shown in Figs. 3, 4 and 5,
 respectively, where $u \equiv \rho/[3(H_0^b)^2]$.
At the early stage with $a < 0.6$, the role of $\rho_f$ is effective and
 $\alpha$ increases with $a$, but at the later stage with $1.0 > a > 0.6$,
  $\Lambda$ is dominant and $\alpha$ decreases slowly after a peak.

The $a$ dependences of $1/u, \beta$ and 
$\alpha$  in the case of  $(\rho/\rho^b)_0 \neq 1$ are also
 found to be similar to those in the case of $(\rho/\rho^b)_0 = 1$,
owing to the above correspondence. Here the $a$ dependence of 
$\alpha$ in the case of $(\rho/\rho^b)_0 \neq 1$ for Model 1 (with the
model parameter (\ref{eq:31})) is  shown in Fig. 6.

Moreover, let us define the time-dependent model parameters $\Omega_M
 (t)$ and $\Omega_M^b (t)$ (representing those in the past) by
\begin{equation}
  \label{eq:38}
\Omega_M (t) \equiv \frac{\rho_T}{3H^2} \quad {\rm and} \quad 
\Omega_M^b (t) \equiv \frac{\rho^b}{3(H^b)^2}.
\end{equation}
Then $\Omega_M = \Omega_M (t_0) $ and $\Omega_M^b = 
\Omega_M^b (t_0) $, and we have the ratio  
\begin{equation}
  \label{eq:39}
\Omega_M (t)/\Omega_M^b (t) = (\rho/\rho^b) (1+\beta)/\alpha^2.
\end{equation}
This ratio tends to $1$ for $a \rightarrow 0$. The $a$ dependence of 
$\Omega_M (t)/\Omega_M^b (t)$ is shown in Fig. 7 for the model
parameter (\ref{eq:28a}).

It is concluded,  therefore, that at the later stage the models with the
 fluctuation energy can have a Hubble constant ($H_0 = 73 \sim
 74 \ {\rm km \ s^{-1} Mpc^{-1}}$) larger than that  ($H_0^b = 67.3 \ 
{\rm km \ s^{-1} Mpc^{-1}}$) in the basic models, while, at the early 
stage with large densities, both models have the same Hubble constants 
(in such a way that $H/H^b  \rightarrow 1$  for $a \rightarrow 0$). 
This shows that the Hubble-constant
 problem\citep{planck1,planck2,h1,h2,h3,h4,decay} can be 
solved by taking the fluctuation energy into account.

\begin{figure}
\caption{\label{fig:3}  The $(1/u - a)$ relation in the case of
 $(\rho/\rho^b)_0 = 1$. The ordinate is $1/u$, where  $u \equiv \rho/[3(H_0^b)^2]$.
 $a$ is the scale-factor and $a = 1$ at the present epoch. }
\centerline{\includegraphics[width=10cm]{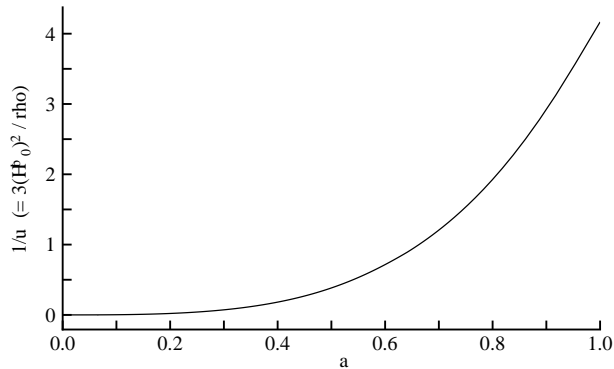}}
\end{figure}
\begin{figure}
\caption{\label{fig:4}  The $(\beta - a)$ relation in the case of
 $(\rho/\rho^b)_0 
= 1$. The ordinate is  $\beta \ (\equiv \rho_f/\rho)$. }
\centerline{\includegraphics[width=10cm]{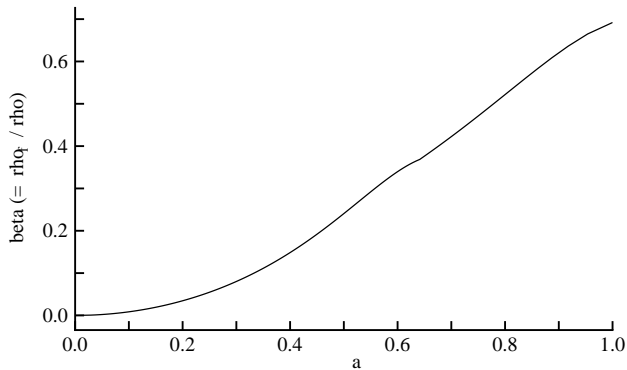}}
\end{figure}
\begin{figure}
\caption{\label{fig:5}  The $(\alpha - a)$ relation
in the case of $(\rho/\rho^b)_0 = 1$. The ordinate is  $\alpha \ (\equiv H/H^b)$. 
$a$ is the scale-factor and $a = 1$ at the present epoch.  }
\centerline{\includegraphics[width=10cm]{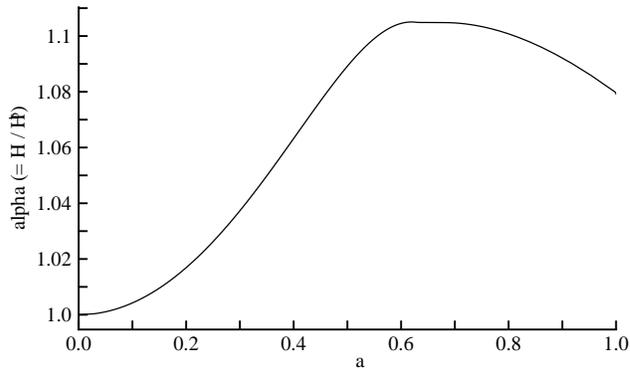}}
\end{figure}
\begin{figure}
\caption{\label{fig:6}  The $(\alpha - a)$ relation in the case of
 $(\rho/\rho^b)_0 = 1.181$. The ordinate is  $\alpha \ (\equiv H/H^b)$. }
\centerline{\includegraphics[width=10cm]{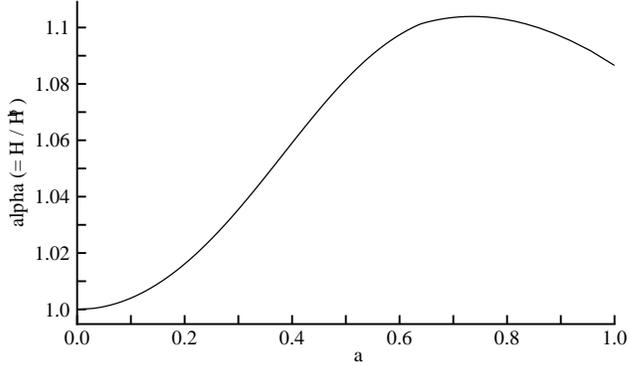}}
\end{figure}
\begin{figure}
\caption{\label{fig:7}  The $(\Omega_M (t)/\Omega_M^b (t) - a)$ 
relation in the case of $(\rho/\rho^b)_0 = 1$. The ordinate is 
$\Omega_M (t)/\Omega_M^b (t) $. }
\centerline{\includegraphics[width=10cm]{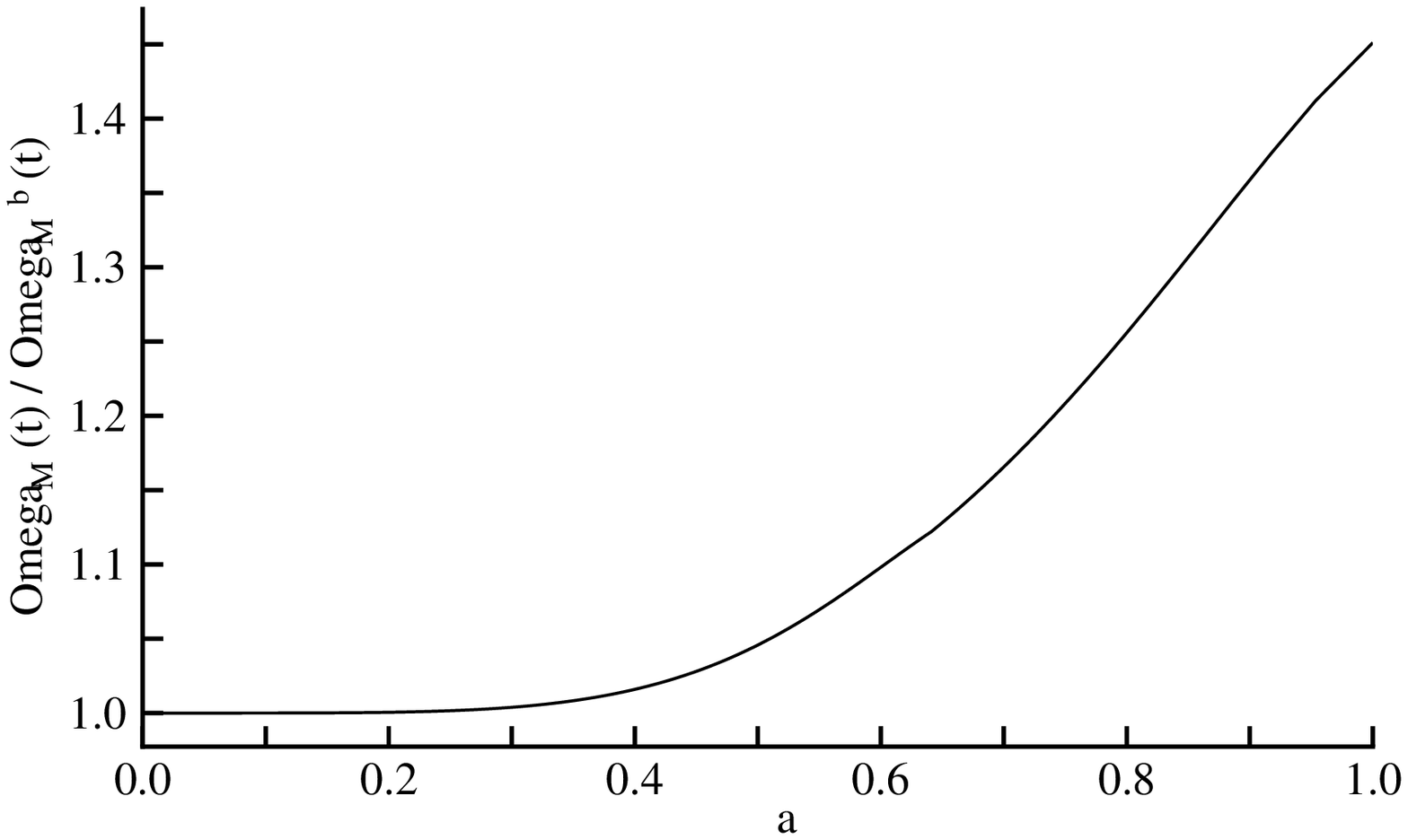}}
\end{figure}
%

\section{Perturbations in cosmological models with the fluctuation energy} 
The behaviors of linear perturbations in the cosmological models with
 pressureless matter are well-known and expressed using the 
gauge-invariant treatment\citep{bar,kodama}. 

Here we assume that the accurate background model has been obtained 
and consider the perturbations to it.
The gauge-invariant density perturbation $\epsilon_T$ for the total density $\rho_T \ (\equiv \rho + \rho_f)$ satisfies the equation
\begin{equation}
  \label{eq:40}
\epsilon_T'' + \frac{a'}{a} \epsilon_T' - \frac{1}{2} (\rho_T a^2) \epsilon_T
= 0.
\end{equation}
The evolutionary relation for the fluctuation energy is assumed to hold 
in the weak inhomogeneities.
Then the gauge-invariant density perturbation $\epsilon$ for ordinary dust
 satisfies 
\begin{equation}
  \label{eq:41}
\epsilon_T = (1+ d \rho_f/d \rho)  \epsilon =  (1+ \beta + \rho d \beta/d 
\rho)  \epsilon,
\end{equation}
and the perturbation $\epsilon_f$ of the fluctuation energy is expressed as
\begin{equation}
  \label{eq:42}
\epsilon_f \equiv \epsilon_T -\epsilon  =  \epsilon_T (\beta + \rho 
d \beta/d \rho) /(1+ \beta + \rho d \beta/d \rho) ,
\end{equation}
and
\begin{equation}
  \label{eq:43}
\epsilon = \epsilon_T/  (1+ \beta + \rho d \beta/d \rho) .
\end{equation}
At the early stage of $\beta \ll 1$, $\epsilon \simeq \epsilon_T$ and 
$\epsilon_f \ll \epsilon$, but at the later stage of $\beta \sim 1$, $\epsilon$ 
and $\epsilon_f$ are comparable.

\section{Concluding remarks}
The existence of random fluctuations is beyond doubt and their amplitudes 
 are also well-known\citep{wein,ll,dodel,tsuji,bbks}. We must take their 
energy (the fluctuation 
energy) into account, to clarify the dynamical evolution of the
universe. This paper is the first step to considering it as a kind of 
 dark matter.

At the stage of $a \ll 1$, the fluctuation energy $\rho_f$ is negligibly small 
compared with the density $\rho$ of ordinary dust, but at the present
epoch it occupies about $36 \sim 41 \%$ of the total density of
the pressureless matter, depending on the basic models.
The fluctuation energy was  considered in this paper as part of 
the dark matter, which 
cannot be touched but contributes to the formation and evolution of 
astronomical objects at the later stage. The essential difference between
the model with the fluctuation energy and the basic models is the
quantitative large change in the dark matter.

In this paper we tentatively adopted Model 1 and Model 2 as the
basic model, to derive
the fluctuation energy using the second-order perturbation theory. 
The derived model parameters depend sensitively on their basic model
 parameters, the present ordinary dust density ratio
 $(\rho/\rho^b)_0$, and the upper limit $x_{max}$ for the integrations 
 $A$ and $B$ (in Appendix A).  Therefore, they  should be selected, so that 
 the derived model parameters may be fitted as well as possible with 
 the observational ones.

In the previous paper\citep{tompre}, we took the effect of fluctuation
 energy into 
account, by renormalizing the model parameters of a basic background 
model due to adding the second-order density and metric perturbations to
the background quantities.  That method is different from the present 
one in which the cosmological models are constructed by taking
the fluctuation energy into account as part of pressureless matter. 
However, we could obtain  similar
model parameters that are consistent with their observational values.

The accuracy for the second-order perturbations $\rho_f \ (\equiv 
\delta_2 \rho)$ is good at the early stage of the universe, because 
$\beta \equiv \rho_f/\rho \ll 1$, but it becomes worse with the
expansion of the universe. At the present epoch, $\beta$ is still 
smaller than $1$, but not so small, i.e. $0.552$ and $0.685$ for the two
basic models as Eq. (\ref{eq:5}) shows.  So, to derive a more accurate model
at the stage of $a \simeq 1$, we should correct $\beta (x)$ in Eq.(\ref{eq:6}) 
- Eq.(\ref{eq:7a}), by constructing the higher-order general-relativistic
 perturbation theories.  

The contributions of the super-sample modes (i.e. the large-scale modes
longer than the survey scales) to the mean density fluctuations and 
the power spectrum in the finite-volume survey have recently been
studied by several authors.\citep{TH, LHT}  They are not equal to
the backreaction of long-wavelength random fluctuations, but they 
may be closely connected with it, and so with the present analyses. If so, 
the general-relativistic second-order perturbations, or the nonlinear 
perturbations in the post-Newtonian approximation may play important roles
also in their treatments (in the similar way to our treatment in the previous
 paper\citep{tompre}). This is because the large-scale modes cross
 the Hubble-scale length during their evolution from the very early
 stage to the present epoch.\citep{baldauf}      
 
 \bigskip
\section*{Acknowledgement}
The author thanks the referee for helpful comments. 
\bigskip

\appendix
\section{Second-order density perturbations corresponding to the 
first-order random fluctuations}
In the Sect. 3 of the previous paper\citep{tompre}, we obtained the
formula for the 
spatial average of the second-order density perturbations in the basic
 models. It is expressed
as
\begin{equation}
  \label{eq:a1}
\langle  \mathop{\delta}_2 \rho/\tilde{\rho} \rangle = \frac{4\pi}{3} \ 
(K_{eq})^4 \
 {\cal P}_{{\cal R}0} \ \frac{[1 - Y(a)]}{(\Omega_M/a + 
 \Omega_\Lambda a^2)} \Big[\frac{11}{2} \ (K_{eq})^{-2} A + Z(a) B\Big], 
\end{equation}
where $ {\cal P}_{{\cal R}0} = 2.2\times 10^{-9}, \ \tilde{\rho} = \rho +
\Lambda, \ H_0^b= 100h,$ and
$K_{eq} \equiv k_{eq}/H_0^b = 219 (\Omega_M^b h).$

For $h = 0.673$, we have
\begin{equation}
  \label{eq:a2}
K_{eq} = 32.4 (\Omega_M^b/0.22).
\end{equation}
For the transfer function $T_s (x)$, $A$ and $B$ are expressed as
\begin{equation}
  \label{eq:a3}
 A \equiv \int^{x_{max}}_{x_{min}} dx \ x \ T_s^2 (x),  \quad   B \equiv
  \int^{x_{max}}_{x_{min}} dx \ x^3 \ T_s^2 (x),
\end{equation}
where $x \equiv k/k_{eq}$ for the wave-number $k$, and the
upper and lower limits of the integrations are specified by
$x_{max} = 5.7$ and $x_{min} = 0.01$.

The definitions of $Y(a),$ and $Z(a)$ are found in the previous
 paper\citep{tompre}.


\end{document}